\documentclass[preprint,letterpaper,amsmath,amssymb,floatfix,nofootinbib]{revtex4}
\usepackage{graphicx,epsfig}
\usepackage{amsmath,amssymb}
\usepackage{epsfig}
\usepackage{subfigure}
\usepackage{color}

\newcommand{\be}{\begin{equation}}
\newcommand{\ee}{\end{equation}}

\begin{document}

\title{Cosmological Problems with Multiple Axion-like
Fields} \author{Katherine J. Mack$^{1,2}$}\email{mack@ast.cam.ac.uk} \author{Paul J. Steinhardt$^{3,4}$}\email{steinh@princeton.edu}
\affiliation{$^1$Kavli Institute for Cosmology, Institute of Astronomy, University of Cambridge, Madingley Road, Cambridge, CB3 0HA, UK} \affiliation{$^2$Department of Astrophysical Sciences, Peyton Hall - Ivy Lane,  Princeton, NJ 08544, USA} \affiliation{$^3$Joseph Henry Laboratories, Princeton University, Princeton, NJ 08544, USA} \affiliation{$^4$Princeton Center for Theoretical Science, Princeton University, Princeton, NJ 08544, USA}

%\date{}

\begin{abstract}

Incorporating the QCD axion and simultaneously satisfying current constraints on the dark matter density and isocurvature fluctuations requires non-minimal fine-tuning of inflationary parameters or the axion misalignment angle (or both) for Peccei-Quinn symmetry-breaking scales $f_a > 10^{12}$ GeV.  To gauge the degree of tuning in models with many axion-like fields at similar symmetry-breaking scales and masses, as may occur in string theoretic models that include a QCD axion, we introduce a figure of merit ${\cal F}$ that measures the fractional volume of allowed parameter space: the product of the slow roll parameter $\epsilon$ and each of the axion misalignment angles, $\theta_0$. For a single axion, $\mathcal{F} \lesssim 10^{-11}$ is needed to avoid conflict with observations.
We show that the fine tuning of $\mathcal{F}$ becomes exponentially more extreme in the case of numerous axion-like fields.  Anthropic arguments are insufficient to explain the fine tuning because the bulk of the anthropically allowed parameter space is observationally ruled out by limits on the cosmic microwave background isocurvature modes.  Therefore, this tuning presents
a challenge to the compatibility of string-theoretic models with light axions and inflationary cosmology.

\end{abstract}

\maketitle

\noindent {\bf Introduction}

The axion is today considered a leading solution to the strong-CP problem as well as a dark matter candidate and a possible feature of string theory.
The axion solution to the strong CP problem \cite{tHooft:1976a,tHooft:1976b} entails the introduction of a global $U(1)$ symmetry, known as the $U(1)_{PQ}$ or Peccei-Quinn (PQ) symmetry \cite{Peccei:1977hh}, that is spontaneously broken to a discrete subgroup $Z_N$, generating a pseudo-Nambu-Goldstone boson, known as the axion \cite{Wilczek:1978,Weinberg:1978}.  Through non-perturbative interactions, the axion, $\phi$, obtains a periodic potential with $N$ distinct minima separated by $\Delta \phi = 2 \pi f_a$ where $f_a$ is the PQ symmetry breaking scale.
In the original proposal, $f_a$ was set to the electroweak scale, $f_a \sim$ TeV, resulting in an axion with large couplings to matter fields; this was quickly ruled out.  It was soon realized, however, that the symmetry-breaking scale could be much higher, reducing the couplings to matter and rendering the axion difficult to detect. This possibility, known as the ``invisible axion,'' is the realization addressed in modern discussions \cite{Kim:1979if,Dine:1981rt}.

In the big bang inflationary model, two scenarios of axion production are possible.  If the symmetry-breaking scale $f_a$ is less than the Hubble parameter $H_I$ at the end of inflation, the axion is non-uniform when it is generated and the universe breaks up into tiny regions separated by domain walls (assuming $N>1$); the domain wall energy dominates the universe, resulting in an unacceptable cosmology.  We refer to such axions as {\it low-$f_a$ axions}.
Alternatively, if PQ symmetry breaking occurs before inflation ({\it high-$f_a$ axions}), the axion field is made uniform throughout the observable universe during inflation and set at some random value, typically displaced from the minimum of its potential.  
The fractional shift from the nearest minimum, $\Delta \phi / (2 \pi f_a)$, known as the {\it misalignment angle}, determines the amplitude of the field once it starts to oscillate about the minimum.  The oscillations of the coherent Bose-Einstein condensate are pressureless, like dust; this non-thermal energy is a candidate for explaining cold dark matter \cite{Preskill:1982cy}.  Numerous experiments have been attempted to detect the axions produced by this ``misalignment mechanism'' (see, e.g., \cite{Asztalos:2001tf,Asztalos:2006kz,Duffy:2009ig}).  
Assuming a value of the misalignment angle $\theta_0 \sim \mathcal{O}(1)$, which is typical, a value of $f_a \lesssim 3 \times 10^{11}$ GeV is required to avoid the overproduction of axionic dark matter via misalignment \cite{Fox:2004kb,Wantz:2009it}, and a tighter bound of $f_a \lesssim 3 \times 10^{10}$ GeV arises when the radiation from axionic strings is taken into account \cite{Wantz:2009it}.  A lower bound also arises from axion couplings to photons and other particles: $f_a \gtrsim 10^{9}$ GeV is required to evade other astrophysical and laboratory constraints (see, e.g., \cite{Duffy:2009ig}).

It is possible to accommodate a high-$f_a$ QCD axion with $f_a \gtrsim 10^{12}$ GeV only if we live in a rare region of space with $\theta_0$ exponentially small compared to unity.  To explain the unlikely value, anthropic selection must be invoked based on the argument that a high axion density (much greater than the observed dark matter density) is hostile to the development of life \cite{Hertzberg:2008wr,Linde:1991km,Tegmark:2004qd,Tegmark:2005dy,Arvanitaki:2009fg}.

In addition to the misalignment, quantum fluctuations induced in the axion field during inflation influence axion field oscillations; even if $\theta_0=0$, inflationary fluctuations displace the field from its minimum and lead to the production of some axion particles.  The rms fluctuation in the axion misalignment angle due to inflationary perturbations is
\be
\sigma_\theta \approx \frac{H_I}{2 \pi f_a} .
\ee
Therefore, to suppress the production of axion particles after inflation, for $f_a > 10^{12}$ GeV, it is not only necessary to have an unlikely misalignment angle, but also the inflationary energy scale must be small.  As shown below, this requires finer tuning of the inflationary slow-roll parameter, $\epsilon$, than is needed to solve the cosmological flatness and horizon problems.

The axion scenario may be realized in string theory.  As pointed out by many authors (e.g., \cite{Svrcek:2006yi,Conlon:2006tq,Arvanitaki:2009fg}), compactifications in string theories always generate PQ-like symmetries and generically produce a multitude of axions and axion-like fields (ALFs).  The number and properties of the ALFs depend on the specific model being considered.
For the case of the heterotic string, there are two classes of ALFs.  The so-called ``model-independent'' axion arises from the antisymmetric tensor field $B_{\mu \nu}$, where $\mu, \nu$ are the usual space-time indices, that plays a key role in anomaly cancellation \cite{Witten:1984dg}.  This axion, which has analogs in other asymptotic limits of string theory and does not depend on compactification, has a symmetry breaking scale $f_a$ of order $10^{16}$ GeV \cite{Polchinski:1998rr,Kim:1999dc,Fox:2004kb,Svrcek:2006yi}.
In the other class of ALFs are the ``model-dependent'' axions whose properties depend upon the compactification of the manifold.  These arise from $B_{ij}$, where $i,j$ are internal indices, and typically have $f_{ALF}$ values between $10^{15}$ and $10^{18}$ GeV.  It is possible to construct models with $f_{ALF}$ somewhat lower (for example, in warped heterotic string theory; see \cite{Dasgupta:2008hb}), though these models are not generic.  
Cosmological constraints apply equally to the QCD axion and to ALFs with high symmetry-breaking scales $f_{ALF} \gtrsim 10^{12}$ GeV.

ALF symmetry-breaking scales $f_{ALF}$ and masses ($m_{ALF} \sim \Lambda_{ALF}^2/f_{ALF}$, where $\Lambda_{ALF}$ is a scale set by the ALF coupling to instantons) may span a large range, with each ALF associated with its own gauge group \cite{Svrcek:2006yi}.  This range can conceivably include ALFs with $\Lambda_{ALF}$ values ranging from the QCD scale to the string scale. 
If all ALFs are very massive ($\Lambda_{ALF} \gtrsim H_I$),
they begin oscillating before inflation, producing particles that are inflated away and are, therefore, harmless.  
However, if string theory can produce a QCD axion with a large symmetry-breaking scale $f_a$ and a very small mass ($< 1$ eV), it is reasonable to expect the spectrum to include other ALFs whose mass and $f_{ALF}$ are within a few orders of magnitude of those of the QCD axion.
Further, as argued for example in Ref.
\cite{Arvanitaki:2009fg,Svrcek:2006yi,Conlon:2006tq}, string theory models could produce {\it many} ALFs with $f_{ALF}$ values near the GUT scale ($> 10^{15}$ GeV).  The masses of these ALFs should be homogeneously distributed on a log scale, with possibly several ALFs per decade of energy \cite{Arvanitaki:2009fg}.
Like the QCD axion, all of these ALFs that begin oscillating after the end of inflation (i.e., those with $f_{ALF} \gtrsim H_I \gtrsim \Lambda_{ALF}$) can contribute significant densities of dark matter, and can produce isocurvature perturbations observable in the cosmic microwave background (CMB).

Previous works have made use of the possibly large number of string theoretic ALFs to construct models of inflation \cite{Dimopoulos:2005ac} and quintessence \cite{Svrcek:2006hf}.  In those constructions, the effect of the ALFs is to contribute to accelerated expansion when the axion field is overdamped by Hubble expansion.  In this work, we consider the limit in which string theoretic ALFs have underdamped oscillations which act as dark matter.  
We show how the existence in string theory of many axion-like fields can exponentially exacerbate the cosmological and fine-tuning problems for axions in the context of inflationary cosmology.

To quantify the fine tuning required to evade cosmological constraints in the presence of multiple ALFs, we introduce a figure of merit $\mathcal{F}$ which measures the fractional volume of allowed parameter space including the tuning of ALF misalignment angles and the slow-roll parameter $\epsilon$ of the inflationary model.  Both $\theta_0$ and $\epsilon$ are $\mathcal{O}(1)$ in an untuned model.  We define:
\be \label{eq:fom}
\mathcal{F} \equiv \epsilon \displaystyle\prod^{n_{ALF}}_{j=1} \theta_0^j 
\ee
where $n_{ALF}$ is the number of ALFs, labeled by $j$, each having its own misalignment angle $\theta_0^j$.  The slow-roll parameter is
\be
\epsilon = \frac{m_{\textrm{Pl}}^2}{2} \left(\frac{V'}{V}\right)^2,
\ee
where $m_{\textrm{Pl}} \approx 2.4 \times 10^{18}$ GeV is the reduced Planck mass. 
This parameter measures the flatness of the inflationary potential $V$ and is related to the inflationary equation of state $w \equiv p/\rho$, with $\epsilon = \frac{3}{2} (1+w)$.

Obtaining the 60 e-folds of inflation required to solve the flatness and horizon problems requires $\epsilon \lesssim 10^{-2}$, a modest fine-tuning.  The amplitude of density fluctuations produced during inflation is $\delta \rho / \rho \sim H_I / (m_{\textrm{Pl}} \sqrt{\epsilon})$.  Since observations require $\delta \rho / \rho \sim 10^{-5}$, we must have $\epsilon = 10^{10} (H_I/m_{\textrm{Pl}})^2$.  This requires $H_I \lesssim 10^{-6} m_{\textrm{Pl}}$; making $H_I$ smaller requires more fine tuning of $\epsilon$ than is necessary for cosmology.  For each order of magnitude reduction in $H_I$, two orders of magnitude additional tuning of $\epsilon$ is required.  (Some have argued that very small $H_I$ is essential to incorporate inflation in string theory; if true, this could be interpreted as a sign that large $H_I$ is ruled out or that inflation and string theory are a poor fit.  However, either interpretation is premature. 
Small $H_I$ models have proven to be problematic \cite{Baumann:2007np} and string theoretic examples with large $H_I$ have been constructed \cite{Silverstein:2008sg}.  So, without a stronger microphysical argument, treating $\epsilon$ as a free parameter with uniform prior is the conservative choice.)
Defining the figure of merit as in equation (\ref{eq:fom}) provides a measure of the fractional volume of parameter space where each $\theta_0$ is independent and stochastic and $\epsilon$ is also independent of each of the ALFs.

A single QCD axion corresponds to fine tuning at the level of $\mathcal{F} \lesssim 10^{-11}$.
In a companion paper \cite{Mack2009}, we show that, in the majority of the parameter space, the tuning is necessary primarily to avoid isocurvature modes.  Since there is no known reason why an observable isocurvature contribution makes the universe uninhabitable, anthropic reasoning cannot explain the required fine tuning.  In order for the dark matter density constraint to dominate the limit on the misalignment angle, one must carefully tune $\epsilon$, which is also unsupported by anthropic considerations.

In this work we will show that each additional ALF with comparable mass exponentially exacerbates the fine tuning $\mathcal{F}$.
The result is then used in a Bayesian analysis to compare models in which many ALFs are produced with those in which they are not.  We also find that, as in the case of the single axion \cite{Mack2009}, anthropic reasoning is ineffective because most of the anthropically allowed parameter space is ruled out observationally due to the isocurvature constraint.

\noindent {\bf Cosmological Constraints on Axions and Axion-Like Fields}

The natural value for $f_a$ (or $f_{ALF}$) in string theory is of order the string scale, $ \gtrsim 10^{15}$ GeV \cite{Svrcek:2006yi,Arvanitaki:2009fg}, corresponding to a high-$f_a$ axion.
The axion field is made uniform and frozen at some misalignment angle $\theta_0$ during inflation, and the field remains frozen until the Hubble scale is of order the axion mass, $m_a$, at temperature $T_{osc}$.  The precise form of the temperature-dependent $m_a(T)$ relies on the relationship between $T_{osc}$ and the scale $\Lambda$ at which instanton effects become important;
two limiting cases for the QCD axion are discussed in \cite{Fox:2004kb}.
For $f_a \lesssim 0.26 (\Lambda/ 200 \textrm{ MeV})^2 m_{\textrm{Pl}}$, $T_{osc}$ is greater than $\Lambda_{QCD}$.  In this regime, the axion mass is temperature dependent because the QCD instanton density is temperature dependent -- this density determines the mass.  For larger $f_a$, $T_{osc} < \Lambda_{QCD}$ so $m_a(T)$ is temperature-independent.  The temperature dependence of a QCD axion is not straightforwardly generalizable to ALFs because it depends upon the details of the gauge couplings.
When calculating the cosmological abundance of ALFs, we treat the mass as being independent of temperature; our conclusions are not sensitive to this assumption.

The total density of the QCD axion plus any additional ALFs is currently observationally constrained by the dark matter density and the fraction of isocurvature perturbations in the CMB.
These constraints can be used to set limits on the axion's fundamental scale ($f_a$ or $f_{ALF}$) and its post-inflation misalignment scale $\theta_0$, as well as on $H_I$.  Detailed derivations of the density and isocurvature constraints can be found in, e.g., \cite{Fox:2004kb,Hertzberg:2008wr}.  Here we briefly describe the origins of the constraints and quote the relevant formulae.  Except for the differences due to the treatment of the temperature-dependent mass, the formulae apply equally well for the QCD axion and for ALFs.

Following \cite{Hertzberg:2008wr}, we parametrize the axion density by the late-time axion energy density per photon, $\xi_a$, defined by:
\be
\xi_a \equiv \frac{\rho_a (T_0)}{n_\gamma (T_0)} = \frac{m_a (T_0)}{m_a (T_{osc})} \frac{\rho_a(T_{osc})}{n_\gamma (T_0)} \frac{s (T_0)}{s (T_{osc})}
\ee
where $s$ is the entropy density, $\rho_a$ is the mass density of axions, $n_\gamma$ is the number density of photons, and $T_0$ refers to the present day.  $\xi_a$ is related to $\Omega_a$, the ratio of the axion density to the critical density, by $\Omega_a h^2 \approx \xi_a / 26$ eV with $h$ the dimensionless Hubble parameter ($h \equiv H_0 / 100 $ km/s/Mpc).  The observational constraint on dark matter $\xi_{\textrm{CDM}} \approx 2.9$ eV places an upper limit on the axion density.

Inflationary fluctuations in the axion field that contribute to the axion particle density also lead to isocurvature perturbations in the CMB.  Limits on the fraction of CMB perturbations that are isocurvature can, therefore, place constraints on an axion field existing during inflation.  To test against this limit, we calculate the ratio of the average power in the isocurvature component to the average total power in CMB temperature fluctuations:
\be
\alpha_a \equiv \frac{\langle (\delta T/T)^2_{iso} \rangle}{\langle (\delta T/T)^2_{tot} \rangle}.
\ee
Current observational constraints from CMB observations limit $\alpha_a < 0.072$ \cite{Komatsu:2008}.
In terms of axion parameters, the density and isocurvature can be written
\begin{eqnarray} \label{eq:xialpha}
\xi_a & \approx & \Lambda (\theta_0^2 + \sigma_\theta^2) G, \\
\alpha_a & \approx & \frac{8}{25} \frac{(\Lambda/\xi_m)^2}{\langle (\delta T/T)^2_{tot}) \rangle} \sigma_\theta^2 (2 \theta_0^2 + \sigma_\theta^2) G^2.
\end{eqnarray}
Here, $\Lambda$ is the instanton scale of the axion field (or ALF), $\theta_0$ is the average misalignment angle within the horizon, and $\xi_m$ is the matter energy density per photon.  We neglect dimensionless factors of order unity that correct for anharmonic effects in the axion potential and uncertainties in the temperature dependence of the axion mass.
The dependence of $\xi_a$ and $\alpha_a$ on $f_a$ (or $f_{ALF})$ is absorbed into the function $G$, which accounts for different cases of the temperature dependence of the axion mass.
We have assumed that the ALF mass is temperature-independent near $T=T_{osc}$, so $G_{ALF} \approx 4.4 ( f_{ALF} / m_{\textrm{Pl}} )^{3/2}$.
In the case of the QCD axion, 
$G_{QCD} \approx 2.8 ( \Lambda_{QCD} / 200 \textrm{ MeV} )^{2/3} ( f_a / m_{\textrm{Pl}} )^{7/6}$ for $f_a \lesssim 0.26 (\Lambda_{QCD} / 200 \textrm{ MeV})^2 m_{\textrm{Pl}}$ and $4.4 ( f_a / m_{\textrm{Pl}} )^{3/2}$ for $f_a \gtrsim 0.26 (\Lambda_{QCD} / 200 \textrm{ MeV})^2 m_{\textrm{Pl}}$ \cite{Hertzberg:2008wr}, where $\Lambda_{QCD} \approx 78$ MeV.  New calculations of the axion mass temperature dependence have recently been published that slightly alter these scalings \cite{Wantz:2009it}, but the small quantitative changes do not affect our result.

\begin{figure}[h!]
\centering
\subfigure[]{%
\epsfig{figure=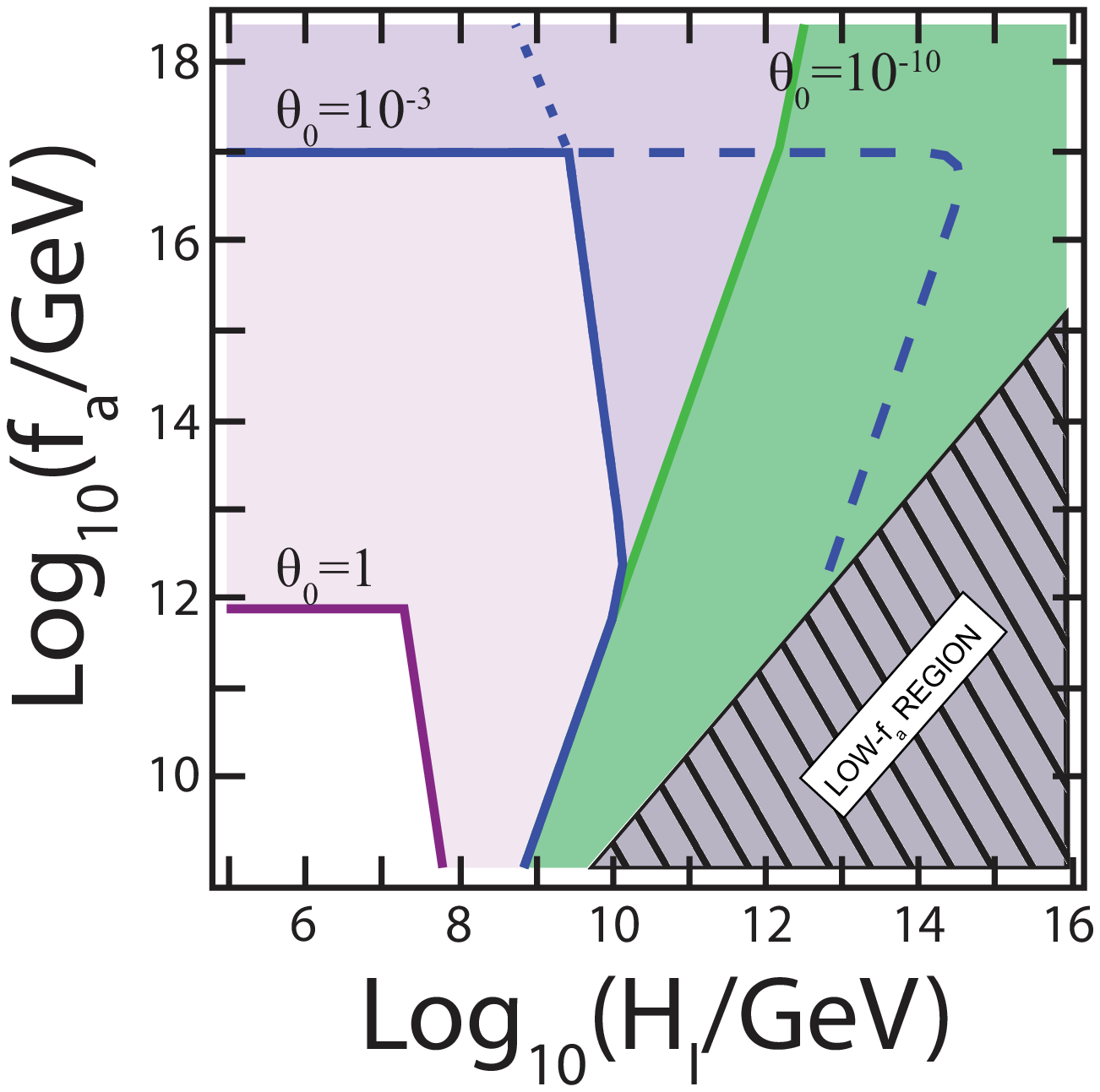,width=3in,angle=0}}\quad
\subfigure[]{%
\epsfig{figure=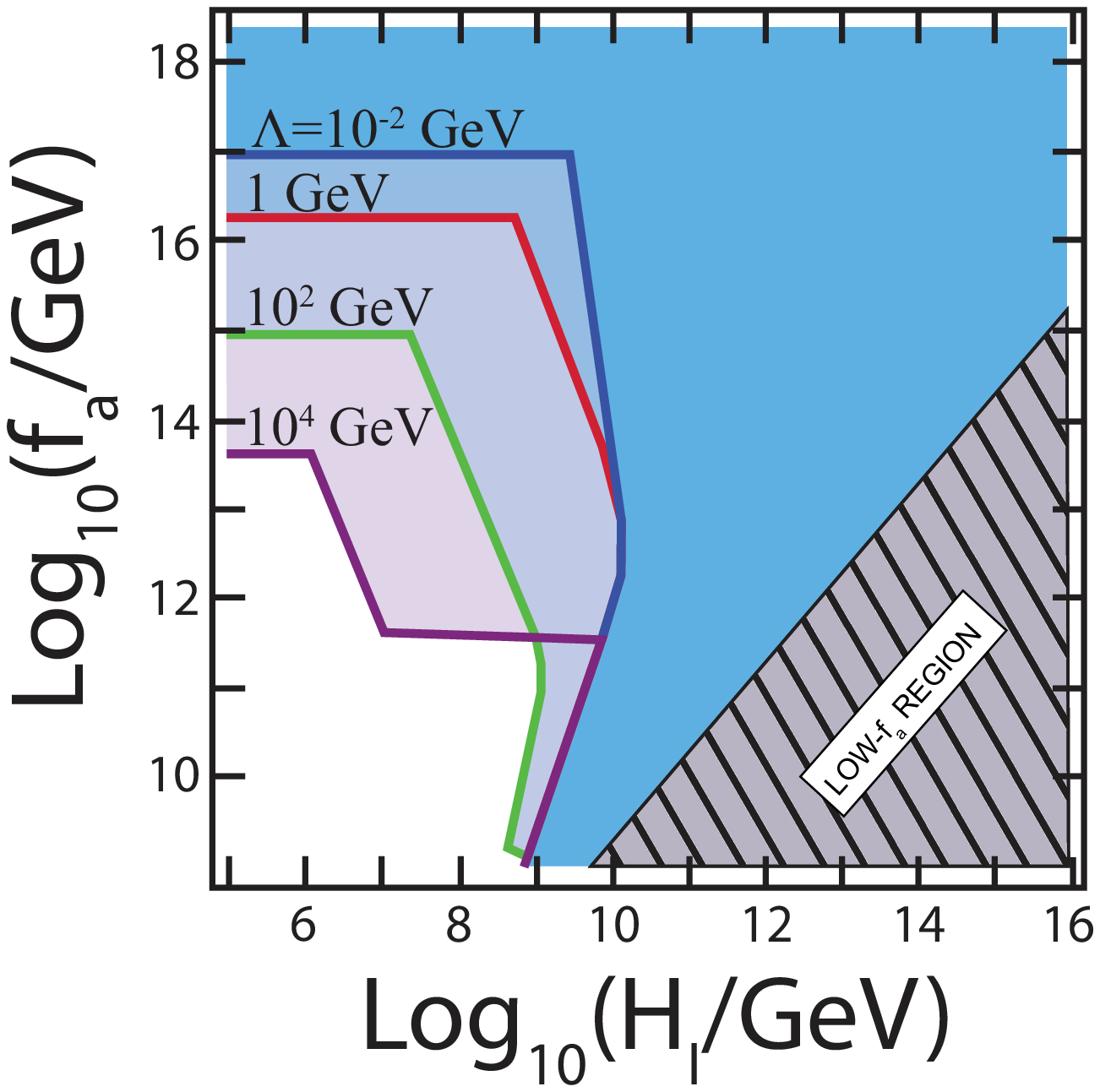,width=3in,angle=0}}
\caption{(a) Cosmological constraints for the standard QCD axion alone for three values of the misalignment angle $\theta_0$.  Each solid curve represents the combined constraint from dark matter density and CMB isocurvature fraction for $\theta_0 = 1, 10^{-3}, 10^{-10}$.  For each solid curve, the shaded region to the right (and/or above) the curve indicates the region ruled out by cosmological constraints.  The dashed and dotted curves indicate for $\theta_0=10^{-3}$ the constraints from dark matter density and CMB isocurvature fraction, respectively. (b) Combined cosmological constraints for the QCD axion plus a single ALF with $\Lambda_{ALF} \neq \Lambda_{QCD}$, and with $\theta_0^{QCD} = \theta_0^{ALF} = 10^{-3}$.  For each solid curve, the shaded region to the right and above is the region ruled out by a combination of the dark matter density and CMB isocurvature constraints.  In both (a) and (b), the white region in the lower left indicates the parameter space that is allowed for $\theta_0 \gtrsim 1$.  The hatched area in the lower right is the low-$f_a$ region ($f_a \lesssim H_I$) not considered in this paper.}%
\label{f:thetalambda}%
\end{figure}

In Figure \ref{f:thetalambda}(a), we plot the cosmological constraints on the QCD axion from measurements of dark matter density and CMB isocurvature for three choices of the misalignment angle: $\theta_0 = 10^{-10}, 10^{-3}$ and $1$.
As can be seen in equation (\ref{eq:xialpha}), the form of the constraints depends strongly upon the relationship between the (unperturbed) post-inflation misalignment angle $\theta_0$ and its rms fluctuation amplitude $\sigma_\theta$ due to inflationary perturbations.  For the density constraint, the presence of the factor $(\theta_0^2 + \sigma_\theta^2)$ implies that, when $\sigma_\theta << \theta_0$, the constraint depends only weakly upon $H_I$ and so primarily sets an upper limit on $f_a$.  In the plot, this produces a horizontal line above which models are ruled out for $\theta_0=10^{-3}$ and 1.
However, when $\sigma_\theta >> \theta_0$ (as for $\theta_0=10^{-10}$ in the figure), the constraint mainly rules out high values of $H_I$.  
For the isocurvature constraint, the dependence on the inflationary perturbation amplitude is stronger, since $\theta_0$ and $\sigma_\theta$ appear in the combination $\sigma_\theta^2(2\theta_0^2 + \sigma_\theta^2)$.  Therefore, the constraint strongly restricts the upper bound on $H_I$, as can be seen in the nearly vertical part of the $\theta_0=10^{-3}$ curve.
When $\theta_0$ is very small ($\theta_0 \lesssim 10^{-4}$), the isocurvature constraint is everywhere more stringent than the density constraint.  For these $\theta_0$ values, the density constraint's cut-off of high $f_a$ values lies above the plot (as in the $\theta_0=10^{-10}$ contour) and its limits on $H_I$ are less stringent than those of the isocurvature constraint.
  As $\theta_0$ increases toward $\mathcal{O}(1)$, the isocurvature constraint is still the tighter constraint unless the Hubble scale of inflation is very low.

%%%%%%%%%%%%%%%%%%%%%%%%%%%%%%%
%%%% Single axion figures %%%%%
%%%%%%%%%%%%%%%%%%%%%%%%%%%%%%%

Dark matter density and isocurvature constraints only apply if the axion or ALF does not decay away in the early universe.
The QCD axion is a long-lived particle.  Its decay into photons 
can occur through an interaction term (e.g., see \cite{Turner:1989vc}) in the Lagrangian: $\mathcal{L}_{int} = g_{a \gamma \gamma} a \mathbf{E} \cdot \mathbf{B} + ...$ where $g_{a \gamma \gamma} \sim \alpha / (\pi f_a/N)$
[up to $\mathcal{O}(1)$ constants], $a$ is the axion field, and $\mathbf{E}$ and $\mathbf{B}$ are the electric and magnetic field vectors, respectively.
The decay rate to two photons
can then be written
$\Gamma_{a \gamma \gamma} \sim g_{a \gamma \gamma}^2 m_a^2 /(64 \pi)$,
with a decay time $\tau = \hbar / \Gamma_{a \gamma \gamma}$.
Comparing this to the age of the universe, $\tau_0 \approx 4.3 \times 10^{17}$ s, we find that the condition for an axion to survive to the present day is:
\be \label{eq:survival}
\left(\frac{f_a}{10^{7} \textrm{ GeV}} \right)^5 \left(\frac{100 \textrm{ MeV}}{\Lambda} \right)^6 \alpha^2 \gtrsim 4.3 \times 10^{-8}.
\ee
This condition is easily satisfied for QCD axions with the symmetry breaking scales we are considering here; their decay into photons can, therefore, be ignored in our calculations.

For ALFs with high instanton scales $\Lambda_{ALF}$, however, decay into gauge bosons may become important.  Using  equation (\ref{eq:survival}) as an estimate leads to an effective upper limit on the scale $\Lambda_{ALF}$ for an ALF to exist at non-negligible densities today.  For the lowest $f_a$ values considered here ($f_a \sim 10^9$ GeV), the upper limit is $\Lambda_{ALF} \lesssim 100$ GeV; at $f_a \sim 10^{16}$ GeV, as expected for general string theory ALFs, the limit is $\Lambda_{ALF} \lesssim 10^8$ GeV.  Hence, we assume $\Lambda_{ALF} < 10^8$ GeV in our analysis.

\noindent {\bf Additional Axion-like Fields}

We now consider the effect of multiple ALFs with masses comparable to that of the QCD axion and $f_a \gtrsim 10^{12}$ GeV on cosmological constraints.  For only a single QCD axion, the free parameters are the misalignment angle $\theta_0$ and the energy scale of inflation.  Each additional axion has its own $\theta_0^i$ and $\Lambda_{ALF}$.

Since the constraints are strongly dependent on the misalignment angle of the axion field, if any one of the ALFs has a large $\theta_0$, it dominates the density and isocurvature contributions.  Similarly, ALFs with larger values of $\Lambda_{ALF}$ at the same $\theta_0$ also dominate.  For the purpose of illustration, we show in Figure \ref{f:thetalambda}(b) the combined constraint (density and isocurvature) for a QCD axion plus one additional axion at a range of $\Lambda_{ALF}$ values, setting the $\theta_0$ values to be the same ($\theta_0^{QCD} = \theta_0^{ALF} = 10^{-3}$).

As $\Lambda_{ALF}/\Lambda_{QCD}$ increases, the ALFs are constrained to lower values of $f_{ALF}$ and $H_I$; recall that $\epsilon$ scales as $H_I^2$.  
In the regime in which $\sigma_\theta << \theta_0$ (i.e., negligible inflationary perturbations), the bound on $f_a$ scales as $\Lambda_{ALF}^{-2/3} \theta_0^{-4/3}$ when $\xi_a$ is held constant or as $\Lambda_{ALF}^{-4/3} \theta_0^{-4/3}$ when $\alpha_a$ is held constant.  In the opposite case where $\sigma_\theta >> \theta_0$, the value of $H_I$ must be lowered to evade the constraints.  The upper limit goes as $H_I \sim \Lambda_{ALF}^{-1/2} f_a^{1/4}$ for $\xi_a$ held constant and $H_I \sim \Lambda_{ALF}^{-1/2} f_{ALF}^{5/8}$ for $\alpha_a$ held constant.

The contributions of axions to the dark matter density and to the isocurvature perturbations are both additive, so each ALF contributes independently to a tightening of those constraints.  If the ALFs are produced with a range of $\Lambda_{ALF}$ values, the constraint is dominated by the subset of ALFs within about an order of magnitude of the largest $\Lambda_{ALF}$.  Among these, the ALFs with the highest $\theta_0$ will provide the dominant contribution to the constraints at the same $f_a$.  Likewise, for models in which the ALFs have a range of $\theta_0$ values, but all else equal, the subset of ALFs with $\theta_0$ values within about an order of magnitude of the highest $\theta_0$ value dominate the constraints, while the contribution of the others is negligible.  Therefore, when comparing actual multiple-ALF models to the constraints and $\mathcal{F}$ values we plot below, one should take $n_{ALF}$ to be the {\it effective} number of ALFs (the number expected to have $\theta_0$ and $\Lambda_{ALF}$ values comparable with the largest value) to compare to the results below.

\begin{figure}[h!]
\centering
\epsfig{figure=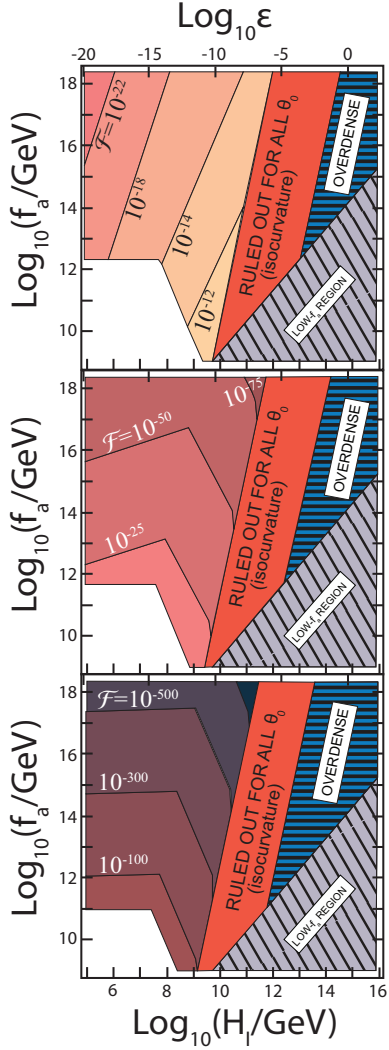,width=2in,angle=0}
\caption{Minimum degree of fine tuning required to satisfy cosmological constraints (as measured by the figure of merit $\mathcal{F}$) for (top) 1, (middle) 10, and (bottom) 100 ALFs with instanton scales of $\Lambda \sim 100$ MeV. The blue hatched region labelled ``overdense''
indicates the region excluded by dark matter density measurements for any value of the misalignment angle $\theta_0$.  For each figure we also plot the region that is ruled out for all values of $\theta_0$ by the CMB isocurvature fraction (red).  In the white region to the lower left, $\theta_0 \gtrsim 1$ (i.e., no tuning of the misalignment angle is necessary).  In the middle and bottom panels, for the purpose of illustration, we assume that each ALF has roughly the same value of $\theta_0$.}%
\label{f:many_axions}%
\end{figure}

In Figure \ref{f:many_axions}, we show exclusion regions from dark matter density and CMB isocurvature constraints and contours of $\mathcal{F}$ for $n_{ALF} = $ 1, 10 and 100 ALFs.  In each plot, the ALFs have $\Lambda_{ALF}=78$ MeV, and all ALFs are assumed to have the same misalignment angle $\theta_0$.  
Keeping all else fixed, the probability of obtaining a certain set of $\{\theta_0^i\}$ values is approximately given by $\prod \theta_0^i$, but the observational constraints depend on $\sum (\theta_0^i)^2$.  Therefore, a set of misalignment angles $\{\theta_0^i\} = \{1, 10^{-10}\}$ is as likely to occur as $\{\theta_0^i\} = \{10^{-5},10^{-5}\}$, but the latter is much less constrained.  Thus models with $\theta_0^i=\theta_0^j$ for all $i,j$ as shown in Figure \ref{f:many_axions} represent the {\it minimum} tuning for a given constraint level.  In the white region in the lower left of the plots, the constraints allow each $\theta_0$ to be of order 1, so the contribution to $\mathcal{F}$ comes only from the tuning of $\epsilon$ (i.e., the inflationary model).  This may be read off the $\log_{10} \epsilon$ axis included along the top of the plots.

We see that for a single ALF (Figure \ref{f:many_axions}, top panel), the least fine tuning (the largest value of $\mathcal{F}$) that can be achieved without violating cosmological constraints is $\sim 10^{-11}$, at low values of the symmetry breaking scale $f_{ALF}$.  This corresponds to keeping $\theta_0$ near its ``natural'' value of order 1 and tuning $H_I$ to be very low ($\sim 10^{9}$ GeV, corresponding to $\epsilon \sim 10^{-11}$).  Alternatively, allowing for tuning of the misalignment angle can be traded for the tuning of $\epsilon$ (e.g., if $\theta_0 = 10^{-9}$ then $H_I$ can be $\sim 10^{12}$ GeV).

For multiple ALFs, the white region in which the values of $\theta_0$ can be $\mathcal{O}(1)$ without violating constraints becomes slightly smaller, while over the rest of the plot area, the parameter $\mathcal{F}$ is constrained to exponentially small values.  This is because, for multiple ALFs, each misalignment angle must be tuned separately to very small values to avoid overproducing isocurvature modes or dark matter particles.  For the 10-ALF and 100-ALF cases (Figure \ref{f:many_axions}, middle and bottom panels), the level of tuning needed is many tens of orders of magnitude worse than the tuning required to solve the strong-CP problem without an axion.  If $\mathcal{F}_{n_{ALF}}$ is the figure of merit for $n_{ALF}$ axions, the degree of tuning scales roughly as $(\mathcal{F}_1)^{n_{ALF}}$.

%%%%%%%%%%%%%%%%%%%%%%%%%%%%%%%%%
%%%% Multiple axion figures %%%%%
%%%%%%%%%%%%%%%%%%%%%%%%%%%%%%%%%

\noindent {\bf Bayesian Analysis}

The exponential fine-tuning required to satisfy observational constraints, as parameterized by $\mathcal{F}$, suggests that the axion explanation of the strong CP problem, inflation, and string theory are not mutually compatible in the absence of some added selection principle.  To quantify the incompatibility, we introduce a Bayesian model comparison, which incorporates both the tuning measure $\mathcal{F}$ for ALFs and the prior probabilities of the paradigms.
A model comparison relies on the calculation of the Bayes factor, which is defined as
\be
B_{M_1,M_0} = \frac{P(D|M_1)P(M_1)}{P(D|M_0)P(M_0)},
\ee
where $P(D|M_i)$ is the posterior probability of the data or observation $D$ in the context of a model $M_i$ and $P(M_i)$ is the prior probability of the model.  A Bayes factor of 3 to 20 is considered to be ``positive'' evidence against model $M_0$, 20 to 150 ``strong'' evidence against, and greater ``very strong'' \cite{Kass:1995}.  In the cases of interest here, $B_{M_1,M_0}$ will be exponentially large.
We use the observed upper limit on the CMB isocurvature mode as the data $D_{iso}$.  The models we compare are:
\begin{itemize}
\item {\bf STwALF:} String theory with one or more light ALFs, assuming a high-$f_a$ QCD axion and inflation,
based on the notion that it is difficult to produce a QCD axion without also producing light ALFs.
\item {\bf no STwALF:} No string theory {\it or} string theory with no light ALFs other than a high-$f_a$ QCD axion, and inflation.
\item {\bf infl:} Inflation, assuming string theory (with ALFs) and a high-$f_a$ QCD axion.
\item {\bf no infl:} An alternative to inflation in which the axion is never excited from its minimum (e.g., the cyclic model \cite{Steinhardt:2006bf}), assuming string theory and a high-$f_a$ QCD axion.
\item {\bf axion:} The high-$f_a$ QCD axion, assuming inflation and string theory (with ALFs).
\item {\bf no axion:} No QCD axion (and an unsolved strong-CP problem), assuming inflation and string theory (with ALFs).
\end{itemize}

We have shown that for a single axion or ALF, the probability of the model agreeing with the data is $\mathcal{F}$.  For the purpose of the Bayesian model comparisons, we are considering the combined effect of ALFs and the QCD axion, so we must also factor in the QCD axion's misalignment angle, $\theta_0^{QCD}$.  Therefore,
\be
P(D_{iso} | \textrm{STwALF + axion + infl}) = \mathcal{F} \times \theta_0^{QCD},
\ee
where $\theta_0^{QCD}$ is the value of the QCD misalignment angle required for it, too, to evade the isocurvature constraint.
Using this, we can calculate Bayes factors expressing the likelihood of each of the three paradigms when the other two are assumed.

For the purpose of illustration, we assume $f_a \sim 10^{16}$ GeV (close to the string scale) for the QCD axion and for ALFs.  The value of $B$ decreases somewhat as $f_a$ decreases, but the qualitative result is unchanged.
A Bayesian model comparison of string theory with light ALFs versus an alternative, assuming the existence of the (high-$f_a$) QCD axion and inflation yields a Bayes factor of
\begin{eqnarray}
B_{(\textrm{no STwALF,STwALF}) | \textrm{infl, axions}} & = & \frac{P(\textrm{no  STwALF})}{P(\textrm{STwALF})} \frac{P(D_{iso} | \textrm{no  STwALF})}{P(D_{iso} | \textrm{STwALF})} \\
  & = & \frac{P(\textrm{no  STwALF})}{P(\textrm{STwALF})} \times \frac{\epsilon^{QCD} \theta_0^{QCD}}{\mathcal{F}} \\
  & = & \frac{P(\textrm{no  STwALF})}{P(\textrm{STwALF})} \times \frac{10^{-12}}{10^{-7} (10^{-4})^{n_{ALF}}} \\
  & \gtrsim & \frac{P(\textrm{no  STwALF})}{P(\textrm{STwALF})} \times (10^4)^{n_{ALF}}
\end{eqnarray}
where $\epsilon^{QCD}$ is the epsilon value required for the QCD axion alone to evade the isocurvature constraint and $10^{-7} (10^{-4})^{n_{ALF}}$ is an approximate expression for the maximum value of $\mathcal{F}$.
As for the ratio of model priors, $P(\textrm{no  STwALF})/P(\textrm{STwALF})$, the current view is that string theory is highly favored compared to alternatives and that string theory with ALFs is more likely than string theory without.  Therefore, this ratio is expected to be less than one, but not sufficiently small to overwhelm the exponential factor $(10^4)^{n_{ALF}}$.  It is one thing to say the theory is strongly favored, but it is another to say the chance of an alternative is one in ten thousand or exponentially worse.
In the equation above, we also have $P(D_{iso} | \textrm{no ST}) = \epsilon^{QCD} \theta_0^{QCD} \sim 10^{-12}$ since, without string theory, ALFs are not produced, so the probability is the $\mathcal{F}$ value we would calculate for a single QCD axion \cite{Mack2009}.

By a similar analysis, we find
\be
B_{(\textrm{no infl,infl}) | \textrm{STwALF, axions}} = \frac{P(\textrm{no  infl})}{P(\textrm{infl})} \frac{P(D_{iso} | \textrm{no  infl})}{P(D_{iso} | \textrm{infl})} \gtrsim \frac{P(\textrm{no  infl})}{P(\textrm{infl})} 10^{12} (10^4)^{n_{ALF}},
\ee
where we take $P(D_{iso} | \textrm{no  infl}) = 1$ because neither $\theta_0$ nor $\epsilon$ are constrained by the CMB isocurvature mode limit;
and
\be
B_{(\textrm{no axion,axion}) | \textrm{infl, STwALF}} = \frac{P(\textrm{no axion})}{P(\textrm{axion})} \times \frac{1}{\mathcal{F} \times \theta_0^{QCD}} \gtrsim 10^2 (10^4)^{n_{ALF}}.
\ee
In this case, we have chosen a model prior that maximally favors the QCD axion as a solution to the strong-CP tuning problem, $P(\textrm{no  axion})/P(\textrm{axion}) \approx 10^{-10}$, since no attractive alternative exists at present.  We have taken $P(D_{iso}|\textrm{axion}) \sim \mathcal{F} \times \theta_0^{QCD}$, where $\theta_0^{QCD} \lesssim 10^{-5}$ is the tuning of the QCD axion misalignment required to evade the isocurvature constraint.
All three Bayes factors, obtained by making conservative estimates for each contribution, become exponentially increasing as $n_{ALF}$ increases.

\noindent {\bf Discussion}

We have considered the fact that string theory models incorporating a QCD axion typically produce additional ALFs with $f_{ALF} \sim 10^{16}$ GeV and masses similar to that of the QCD axion.  
Although string theory may allow some vacua with no additional light ALFs, they are atypical and not favored by any microphysical arguments. We have shown that the multi-ALF models inevitably require a fine tuning of misalignment angles and the inflationary slow-roll parameter $\epsilon$
 that is many orders of magnitude worse than that faced by the QCD axion alone, as measured by the figure of merit ($\mathcal{F}$).  In fact, since the ALF misalignment angles must each be independently tuned to very low values simultaneously to evade cosmological constraints, the tuning is roughly exponential in the number of additional fields.  In addition to the tuning in the ALF values of $\theta_0$, the effect of inflationary perturbations on ALF particle production requires that $H_I$ and hence $\epsilon$ be tuned to a much smaller value than is necessary to solve the cosmological flatness and horizon problems.  Our Bayesian analysis quantifies the conclusion that, absent some strong selection principle, inflation, the QCD axion explanation for the strong-CP problem, and string theory (as currently understood) are not mutually compatible.   

Anthropic reasoning has been invoked in the past as a possible selection principle.  However, as emphasized in \cite{Mack2009} for the case of the QCD axion alone (no additional ALFs), the anthropic principle does not alleviate the problem because most of the anthropically allowed range of parameters is ruled out observationally by the isocurvature constraint.  That is, even after restricting parameters to the range compatible with habitability (based on the density constraint), additional tuning is then required to prevent the generation of isocurvature perturbations whose amplitudes are inconsistent with current measurements of the CMB but consistent with habitability.  This additional tuning is of a magnitude comparable to that of the strong-CP problem the axion was invented to solve.

The same applies to the case of multiple ALFs.  For example, for $f_a \sim 10^{16}$~GeV, as suggested by string theory, the maximum value of $\mathcal{F}$ in the parameter space that is anthropically allowed but ruled out by the isocurvature constraint is much larger than the maximum $\mathcal{F}$ in the observationally allowed space.  Moreover, in some versions of the anthropic principle, it is argued that any anthropically selected variables (such as the dark matter density) should have values near the maximal consistent with
human existence, also known as the anthropic boundary \cite{Weinberg:1987dv,Hall:2007ja}.
If the axionic dark matter density is taken to be near the anthropic boundary, large isocurvature amplitudes are strongly preferred, which is at odds with observation.

Since anthropic reasoning does not force the isocurvature mode fraction or the energy scale of inflation to be in the narrow range compatible with observations, a compelling argument from fundamental physics is needed.  Alternatively, at least one of the three ideas -- the axion solution to the strong CP problem, inflationary cosmology, or string theory -- must be abandoned.

%%%%%%%%%%%%%%%%%%%%%%%%%%%%%%%%%%%%%%%%%%%%%%%%%%%%%%

{\bf Acknowledgments:} 
We thank Jeremy Goodman, Daniel Grin, David Spergel, Daniel Wesley and Edward Witten for comments on an early draft of the paper, and Jacob Bourjaily, Fiona Burnell, Joseph Conlon and Chris Hirata for helpful conversations.

\end{document}